\documentclass[11pt]{article}
\usepackage{amssymb,epsfig,theorem,cite}
\newcommand{\nc}{\newcommand}
\nc{\be}{\begin{equation}}
\nc{\ee}{\end{equation}}
\nc{\bea}{\begin{eqnarray}}
\nc{\eea}{\end{eqnarray}}
\nc{\disp}{\displaystyle}
\nc{\ade}{\mbox{$A$-$D$-$E$}}
\nc{\calN}{{\cal N}}
\nc{\calC}{{\cal C}}
\nc{\calM}{{\cal M}}
\nc{\calS}{{\cal S}}
\nc{\phit}{\hat{\varphi}}
\nc{\chit}{\hat{\chi}}
\nc{\hcalN}{\hat{\calN}}
\nc{\hcalS}{\hat{\calS}}
\nc{\hS}{\hat{S}}
\nc{\sigmad}{\sigma^\dagger}
\nc{\psid}{\psi^\dagger}

\def\sstyle{\scriptstyle}
\def\dps{\displaystyle}

\newtheorem{lemma}{Lemma}

\font\tenmsb=msbm10
\font\sevenmsb=msbm7
\font\fivemsb=msbm5
\newfam\msbfam
\textfont\msbfam=\tenmsb
\scriptfont\msbfam=\sevenmsb
\scriptscriptfont\msbfam=\fivemsb

\def\binom#1#2{{#1\choose #2}}

\def\bra#1{\langle #1|}
\def\ket#1{|#1\rangle}
\def\e{{\rm e}}
\def\d{{\rm d}}

\begin{document}

\title{Nonequilibrium stationary states and equilibrium models with long range
interactions
}

\author{Richard Brak$^1$, Jan de Gier$^1$ and Vladimir Rittenberg$^2$
\\[5mm]
{\small \it $^1$Department of Mathematics and Statistics, University
of Melbourne, Parkville 3010, Australia}\\
{\small \it$^2$Physikalisches Institut, Bonn University, 53115 Bonn, Germany.
}}

\date{\today}
\maketitle
\begin{abstract}
It was recently suggested by Blythe and Evans
that a properly defined steady state normalisation factor can be seen as a
partition function of a fictitious statistical ensemble in which the
transition rates of the stochastic process play the
role of fugacities. In analogy with the Lee-Yang description of phase
transition of equilibrium systems, they studied the zeroes in the complex
plane of the normalisation factor in order to find phase transitions in
nonequilibrium steady states. We show that like for equilibrium systems,
the ``densities'' associated to the rates are non-decreasing functions of
the rates and therefore one can obtain the location and nature of phase
transitions directly from the analytical properties of the ``densities''. We
illustrate this phenomenon for the asymmetric exclusion process. We
actually show that its normalisation factor coincides with an
equilibrium partition function of a walk model in which the ``densities'' have
a simple physical interpretation.
\end{abstract}

\section{Introduction}
The extension of concepts used in equilibrium statistical mechanics, like
the free energy, to nonequilibrium steady states has a long history
\cite{DerrLS01,BazhLZ99}. That a simple extension is not possible can be seen
in \cite{ArndtHR98} where it was shown that in certain cases the free energy
functional is not a convex function of the density. On the other hand
Arndt \cite{Arndt00} has shown in an example that applying the Lee-Yang description
using the zeros of an ad hoc definition of a grand-canonical partition
function gives the correct phase transition. Further applications
of this idea can be found in \cite{ArndtDH01,DammerDH01,BlytheE02}.
In a very interesting new development Blythe and Evans
\cite{BlytheE03} considered the normalisation of the
stationary state of several stochastic systems as a function of the
transition rates and applied the Lee-Yang approach in the same way as one
would for an equilibrium partition function. While a normalisation may seem
to be defined ambiguously, it was noted in \cite{Blythe01,BlytheE01} that
in a formal way, a unique definition of this normalisation can be made
using the matrix-tree theorem, which has a long history in graph
theory going back to Sylvester \cite{Sylvester55}, see
also \cite{Chen76,Chaiken82} and Section \ref{se:matrixtree}. This
connection explicitly relates the normalisation of a stationary state
to the combinatorial problem of counting weighted spanning trees on
graphs, which implies a direct interpretation of the normalisation as
a statistical mechanical partition sum. Can we can learn anything
about the steady state phase diagram from the spanning trees? There
are a few problems with that. For example, it is generically unclear
what the correspondence is between original dynamical quantities and
those that are natural for describing the spanning trees. Secondly,
the normalisation as defined 
above may not be ``minimial'' in the sense that it may contain an
overall nontrivial polynomial factor which is common to each of the
stationary state weights. Such a polynomial factor cannot contribute
to the phase behaviour of the stationary state. We will call the
normalisation as defined via the matrix-tree theorem but with common
factors removed the {\em reduced} normalisation.

The purpose of this paper is to try to bring a better understanding of the
Blythe-Evans approach which is summarized in Section
\ref{se:matrixtree}. In this section we also show that to each 
transition rate one can formally associate 
``particle numbers'' the same way one relates particle numbers to
fugacities. Moreover, like in thermodynamics, one can prove that the
``particle numbers'' are non-decreasing functions of the transition rates.
This important observation allows us to detect the existence of phase
transitions from the behavior of the ``particle numbers'' in the space of
transition rates. At this point the physical meaning of the ``particles'' is
completely obscure (there are as many ``particles'' as the number of independent
transition rates minus one). Moreover, the thermodynamic potential
defined through the (reduced) normalisation factor is not an extensive
quantity. The volume, defined by the leading asymptotic behaviour of
the normalisation, and therefore the definition of the ``densities''
might change in the space of transition rates. This allows for phase
transitions not encountered in equilibrium models with local
interactions. This phenomenon appears in the following way. In
a certain domain of the fugacities, the ``densities'' span the entire
interval between zero and one. This defines a ``phase'' (inside this
domain one can have, like in equilibrium, several phase
transitions). The boundary of the domain separates it from another
domain (``phase'') where one has to take another definition of the
``densities'' because of the change of the volume. In this second phase
the ``densities'' are not necessarily finite.

The fact that the reduced normalisation factor might have a direct physical
interpretation is known from the raise and peel model
\cite{GierNPR03}. This is a one-dimensional stochastic model with
nonlocal transition rates and its reduced normalisation factor (whose
logarithm is proportional to the square of the system size) coincides 
with the number of configurations of the two-dimensional ice model with
domain walls boundary condition $-$ an equilibrium problem. It is our aim to
show that this situation is more general.

In Section~{\ref{se:otw} we define the one-transit walk model (OTW). This model,
which is not parity invariant, depends on two parameters which are
the Boltzmann weights or fugacitites of contact points. We
compute the partition function of this model as well as the two densities
corresponding to the two fugacities. The two densities have a clear physical
meaning. The phase diagram of the OTW model is obtained from the
expressions for these densities. It is the same as that of the totally
asymmetric simple exclusion process (TASEP)
\cite{DerrDM92,DerrEHP93,SchuetzD93,Derrida98,Schuetz00} if we replace
the two fugacities with the boundary rates of TASEP. As we are going to show, the derivation
of the phase diagram of the TASEP model from the analytic
properties of the number of ``particles'' as a function of
fugacities will give a better understanding of the nature of the phase
transitions. In Section \ref{se:rel} we discuss the microscopic properties of the
OTW model making cleare the connection with the TASEP. We conclude with a
discussion of the phase diagram obtained from the ``densities'' of the
partially asymmetric simple exclusion process (PASEP)
\cite{Sasamoto99,BlytheECE00}. In this case we have three kind of
``particle'' numbers: two associated with the boundary rates and one
associated to the back hopping rate $q$. At the symmetric point $q=1$
a new kind of phase transition occurs.

Our conclusions are presented in Section \ref{se:conc}.

\section{The normalisation as a positive polynomial}
\label{se:matrixtree}
Let us start by considering an arbitrary Markov process in continuous
time on a state space spanned by the states $\{\ket a\}_{a=1}^n$,
whose master equation is given by 
\begin{equation}
\frac{\d}{\d t} \bar{P}_t(a) = \sum_{b\neq a} \left(r_{ab} \bar{P}_t(b) - r_{ba}
\bar{P}_t(a)\right).
\label{eq:master}
\end{equation}
The $r_{ab}$ are the transition rates from state $\ket b$ to $\ket
a$ and $\bar{P}_t(a)$ is the (unnormalized) probability to find the system 
at time $t$ in state $\ket a$. Equation (\ref{eq:master}) can be
conveniently rewritten as
\begin{equation}
\frac{\d}{\d t} \ket{\bar{P}_t} = - H \ket{\bar{P}_t},\qquad \ket{\bar{P}_t} =
\sum_{a=1}^n \bar{P}_t(a) \ket a,
\end{equation}
where $H$ is the matrix with off-diagonal elements $H_{ab}=-r_{ab}$
and whose columns add up to zero. One of the main properties of
interest of such a Markov process is its long time behaviour. In the limit $t\rightarrow
\infty$ the system approaches its stationary state
$\ket{\bar{P}_\infty}$, which we will assume to exist and for simplicty to be
unique, given by 
\begin{equation}
H \ket{\bar{P}_\infty} =0.
\label{eq:stationary}
\end{equation}
The stationary state is thus given by the right eigenvector
of the matrix $H$ corresponding to its eigenvalue $0$. This equation
can be solved in the following formal way, see
e.g. \cite{Seneta81}. Let $H(a,b)$ be the matrix corresponding to $H$
with the $a$th row and $b$th column removed. The 
cofactor $X(a,b)$ is then defined by, 
\begin{equation}
X(a,b) = (-1)^{a+b} \det H(a,b).
\end{equation}
If the eigenvalue $0$ is unique,
\begin{equation}
0 = \det H = \sum_b H_{ab} X(a,b) = \sum_b H_{ab} X(b,b),
\end{equation}
where we have used $X(a,b) = X(b,b)$ for all $a$ (see Appendix
\ref{ap:proof}). We see that the eigenvalue equation
(\ref{eq:stationary}) is solved by the cofactors of $H$, 
\begin{equation}
H \ket{P} =0,\qquad P(b) = X(b,b).
\end{equation}
This solution fixes a particular normalisation of the
eigenvector for all system sizes. This normalisation is uniquely defined 
up to an overall rescaling of $H$, or equivalently a rescaling of time (which can vary 
with the system size). To be able to interpret $P(b)$ as a probability
distribution, we write
\begin{equation}
\bar{P}_\infty(b) = P(b)/Z_n,\qquad Z_n = \sum_{b=1}^n P(b) =
\sum_{b=1}^n X(b,b).
\label{eq:norm}
\end{equation}

It can be shown using the matrix-tree theorem \cite{Chen76,Chaiken82}
that the normalisation $Z_n$ of a stationary state of any stochastic (Markov)
process is always given by a homogeneous polynomial in the rates
$r_{ab}$ (some of which might be equal
or be zero), of degree $n-1$ and with positive coefficients, i.e. it has the
form of a generating function. A simple proof of this important statement 
is for example given in \cite{Seneta81}, which we have included in Appendix
\ref{ap:proof}.

We would like to identify the rates $r_{ab}$ as generalized Boltzmann
factors or fugacities $r_{ab} =z_{ab}$ , and $Z_n(\{z_{ab}\})$ as a 
generalized partition sum for nonequilibrium systems. Since the
normalisation $Z_n$ is a polynomial in the variables $z_{ab}$ with
positive coefficients, by the Cauchy-Schwartz inequality 
its negative logarithm,
\begin{equation}
F_n = -\log Z_n,
\label{eq:Fdef}
\end{equation}
is therefore a convex function in all its arguments $z_{ab}$. In analogy
with equilibrium statistical mechanics, we will associate to each rate
$r_{ab}=z_{ab}$ a ``particle number'' $N_{ab}$,
\begin{equation}
N_{ab}= -z_{ab} \frac{\partial F_n}{\partial z_{ab}}.
\end{equation}
These numbers are positive and increasing functions of the fugacities
for any size of the system but they are linearly dependent. One can
arbitrarily choose one rate equal to one which fixes the time scale and
leaves the remaining rates dimensionless. In this way we are left
with one fugacity less. In the large $n$ limit,
\begin{equation}
N_{ab} = V(n) \rho_{ab},
\label{eq:densdef}
\end{equation}
where $V(n)$ is the volume and $\rho_{ab}$ are the ``densities''. One can now use
the equilibrium approach to the theory of phase transitions and apply it
to the densities $\rho_{ab}$. A first order phase transition, for
example, is a location in the transition rates space where one of the
``densities'' has a discontinuity. 

It may happen however, as it will in our example below, that all
cofactors $X(b,b)$ contain a common nontrival polynomial factor. Such
a common factor will cancel out in $\bar{P}_\infty(b)$ and hence cannot
contribute to the nonequilibrium phase behaviour. In (\ref{eq:Fdef})
however it could give rise to spurious singularities that are not
related to the physical phase transitions. In the example of Section
\ref{se:otw} no such spurious phase transitions appear (see Section
\ref{se:tasep}).  

There is another major difference however between equilibrium
systems with short-range interactions and the present problem: the
``particle numbers'' are not necessarily extensive quantities (see the
examples in Sections \ref{se:thermo} and \ref{se:pasep}). This implies
that in the parameter space the $\rho_{ab}$ might diverge and we have to
change the definition of the factor $V(n)$ in
(\ref{eq:densdef}). Actually such a phenomenon is also known in
equilibrium problems with non-local interactions (see \cite{BrakEO98})
in the theory of special surface phase transitions \cite{Binder83}. As
we are going to show in Section \ref{se:pasep} the analogy goes deeper. 

The philosophy we adopt in this paper, is to assign a physical meaning
to the purely formally defined normalisation factor and ``densities'' 
by looking at simple weighted walk problems
for which we can compute the partition functions. The weights of the
configurations depend on parameters which correspond to the rates of the
stochastic processes and the partition function coincides with the
normalisation factor defined in (\ref{eq:norm}) if a common
factor to all the cofactors is removed. In the next sections we
illustrate this approach with the help of an example. We first
consider a combinatorial problem which is interesting on its own. This
is the one-transit walk model. We will compute its partition function
and obtain the phase diagram of the model from the properties of the
densities. We also show that the same partition function coincides
with the normalisation factor of the TASEP model.
\section{The OTW model versus the TASEP}
\label{se:otw}

The totally asymmetric simple exclusion process (TASEP) has grown to be one of the
main theoretical models of nonequilibrium statistical physics. This is
not only due to its simplicity and general applicablity, but also
because its stationary probability distribution (SPDF) and other
properties can be calculated exactly
\cite{DerrDM92,DerrEHP93,SchuetzD93,Derrida98,Schuetz00}. In this paper we show
that this SPDF can be regarded as an 
equilibrium probability distribution of a simple model of a walk near an
interface. Since the walk model is an equilibrium system, it can be
described thermodynamically using standard methods. The phase behaviour of the TASEP
can be explained in terms of adsorption transitions of the walk on the
interface.  

In Section~\ref{se:walk} we introduce a model of a walk in the vicinity of a fixed
interface. The walk is allowed to penetrate the interface once. Both ends
of the walk are fixed but the point of penetration is free. An excess
interface fugacity $1/z_1$ is associated for contacts above the interface, and a fugacity
$1/z_2$ for contacts below. For this model we are able to calculate the
phase diagram exactly. The walk model without penetration was used as a simple
model for polymer adsorption in \cite{BrakEO98}. After 
discussing the thermodynamics of the walk model, we show in
Section~\ref{se:TASEP} that it is closely related to the TASEP with open
boundaries. More precisely, the statistical partition function
$Z(z_1,z_2)$ of the walk model is equal to the reduced normalisation of the
stationary state of the TASEP if the interface fugacities in the walk
model are equal to the in- and output rates of the two reservoirs.   

We show that the thermodynamic TASEP density $\rho$ and current $J$ are
related to the contact densities $\rho_1$ and $\rho_2$,
conjugate to $z_1$ and $z_2$ respectively, as
\begin{equation}
\frac{2\rho -1}{J} = \frac{1-\rho_2}{z_2}-\frac{1-\rho_1}{z_1},
\end{equation}
where
\begin{equation}
\omega = -\lim_{n\rightarrow \infty}\frac{1}{n}\log Z_n(z_1,z_2), \quad
\rho_i=-z_i \frac{\partial \omega}{\partial z_i}.
\end{equation}
These equations show that one may derive the thermodynamic behaviour
of quantities for a nonequilibrium model from those of an equilibrium
model. We hope to be able to make contact between our approach and
the formulation of a free energy functional for the TASEP from large
deviation functions as adopted in \cite{DerrLS02a,DerrLS03}. We also
would like to point out that the walk model has an appealing analogy
with a continuous model for the dynamics of shocks in terms of which
the TASEP phase diagram can be explained quantitatively
\cite{KolomSKS98}.

\subsection{The one-transit model}
\label{se:walk}

Consider a statistical model of a restricted solid-on-solid (RSOS) path,
also called Dyck path, on the rotated square lattice. Paths start at
$(0,0)$ and end at $(2n,0)$, can only move in the North-East (NE) or
in the South-East (SE) direction and cross the x-axis exactly once,
see Figure \ref{fig:path}. 
\begin{figure}[h]
\centerline{
\begin{picture}(280,110)
\put(0,40){\epsfxsize=300pt\epsfbox{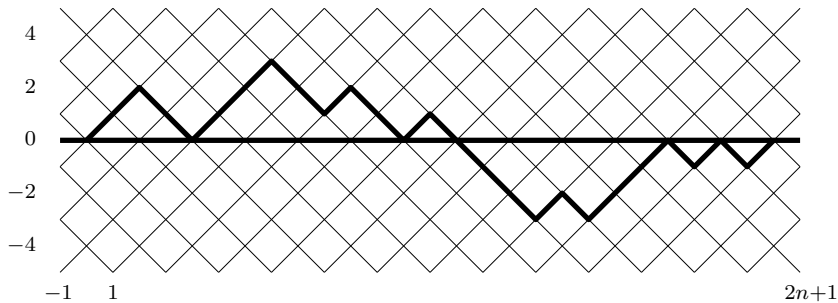}}
\put(0,98){$\sstyle \hphantom{{}-{}} 4$}
\put(0,78){$\sstyle \hphantom{{}-{}} 2$}
\put(0,58){$\sstyle \hphantom{{}-{}} 0$}
\put(0,38){$\sstyle -2$}
\put(0,18){$\sstyle -4$}
\put(14,0){$\sstyle -1$}
\put(38,0){$\sstyle 1$}
\put(294,0){$\sstyle 2n+1$}
\end{picture}}
\caption{An example of an RSOS path starting at $(0,0)$ and ending at $(2n,0)$
crossing the x-axis only once. }
\label{fig:path}
\end{figure}
We associate energies $-\varepsilon_1$ and $-\varepsilon_2$ to the
returns (or contact points) of the path above and
below the x-axis respectively. To make contact with Section
\ref{se:matrixtree} we implement this in the following way. A fugacity
$z_1=\e^{\varepsilon_1/kT}$ is given to each down step, and
$z_2=\e^{\varepsilon_2/kT}$ to each up step, except those ending on
the x-axis. This model is directly related to the canonical model of
\cite{BrakE03}. 

By reflecting the last part of the RSOS path in the x-axis,
it can be easily seen that the total number of possible paths is equal
to the number of Dyck paths of length $2n$. It is known, see
e.g. \cite{BrakE01} that the number of Dyck paths with $p$ returns is
given by Ballot numbers,
\begin{equation}
B_{n,p} = \frac{p}{n}\binom{2n-p-1}{n-1} =
\frac{p(2n-p-1)!}{n!(n-p)!}.
\label{eq:Dyckreturn}
\end{equation} 
The total number $C_n$ of Dyck paths of length $2n$ can be obtained by
summing over $p$ in (\ref{eq:Dyckreturn}), or by noting that it is
equal to the number of Dyck paths of length $2n+2$ with exactly one
return,
\begin{equation}
C_n = \sum_{p=1}^{n} B_{n,p} = B_{n+1,1} = \frac{1}{n+1}
\binom{2n}{n}, 
\label{eq:catalan}
\end{equation}
which is the Catalan number. The partition function of the one-transit
model is simply given by
\begin{equation}
Z_{n}(z_1,z_2) = (z_1z_2)^n \tilde{Z}_n(z_1,z_2),
\end{equation}
where
\begin{equation}
\tilde{Z}_n(z_1,z_2) = \sum_{p=0}^n B_{n,p} \sum_{q=0}^p z_1^{-q} z_2^{-p+q}.
\label{eq:partsum1}
\end{equation}
This can also be written in the following way,
\begin{eqnarray}
Z_{n}(z_1,z_2) 
&=& (z_1z_2)^n\sum_{p=0}^n \tilde{Z}_{p}(z_1,\infty) \tilde{Z}_{n-p}(\infty,z_2).
%
\label{eq:partsum2}
\end{eqnarray}
This formula shows that we can also interpret
our model as the combination of two contact models with a movable but 
impenetrable wall in between them at a random position, each position
being equally probable. Equation (\ref{eq:partsum2}) thus defines the
partition function of an annealed system, i.e. where the partition sum
is averaged over the random position of the wall. 

The partition sum (\ref{eq:partsum2}) is equal to the reduced
normalisation of the totally asymmetric exclusion simple process
(TASEP) \cite{DerrDM92,DerrEHP93,SchuetzD93,Derrida98,Schuetz00}
if the fugacities $z_1$ and $z_2$ are replaced by its boundary
in- and output rates. This result is important since it will allow us
to associate the ``densities'' defined formally from the reduced
normalization of the TASEP with the physical densities of the OTW model.
In Section \ref{se:TASEP} we will go deeper into the
relation between the OTW model and the TASEP, but before that we first
describe the phase diagram of the OTW model.

\subsection{The phase diagram of the OTW model}
\label{se:thermo}
We define the grand potential per site for the gas of contacts as
\begin{equation}
\omega = - \lim_{n\rightarrow \infty} \frac{1}{n} \log Z_{n}.
\end{equation}
Note that to get rid of spurious factors of $2$ in subsequent formulas
we divide by $n$ instead of the system size is $2n$. The potential
$\omega$ can be easily calculated from (\ref{eq:partsum2}) once we
know the asymptotic properties of
$\tilde{Z}_{n}(z,\infty)=\tilde{Z}_{n}(\infty,z)$. This is well known,
see e.g. \cite{DerrEHP93,BrakEO98}, and can for example be derived from
the differential equation it satisfies, 
\begin{equation} 
-(1-z)(1-2z) \tilde{Z}_{n}'(z,\infty) +
 z(z+n(1-2z)^2)\tilde{Z}_{n}(z,\infty) = 2z^2 \frac{(2n-1)!}{n!^2}. 
\end{equation}
Analyzing the large $n$ behaviour of this equation for the regions $z>1/2$,
$z=1/2$ and $z<1/2$ we immediately obtain
\begin{equation}
\tilde{Z}_{n}(z,\infty) \approx
\left\{
\renewcommand{\arraystretch}{2.2}
\begin{array}{cc}
\dps \frac{z}{(1-2z)^2} \frac{4^n}{\sqrt{\pi} n^{3/2}} & z > 1/2 \\
\dps \frac{4^n}{\sqrt{\pi} n^{1/2}}& z=1/2 \\
\dps \frac{1-2z}{1-z}\frac{1}{z^n(1-z)^n} & z<1/2.
\end{array}
\right. 
\label{eq:asymp}
\end{equation}
The grand potential $\omega$ is given by minimizing over the position
of the domain wall, and is therefore given by
\begin{equation}
\omega(z_1,z_2) = -\log 4z_1z_2 + \inf_{0\leq x\leq 1} \,\omega_x (z_1,z_2),
\label{eq:fe}
\end{equation}
where 
\begin{equation}
\omega_x(z_1,z_2) = 
\left\{
\begin{array}{cc}
0 & z_1,z_2 \geq 1/2 \\
\dps x \log 4z_1(1-z_1) +(1-x) \log 4z_2(1-z_2) &  {\rm elsewhere}
\end{array}
\right. 
\end{equation}
From (\ref{eq:fe}) one finds the grand potential in all regions of the
phase diagram,
\begin{equation}
\omega(z_1,z_2) = \left\{
\renewcommand{\arraystretch}{1.6}
\begin{array}{ll}
-\log 4z_1z_2 & \quad \dps z_1,z_2 \geq 1/2 \\
\dps -\log z_2+\log (1-z_1) & \quad\dps z_1 <1/2,\;\; z_2>z_1\\
\dps -\log z_1+\log (1-z_2) & \quad\dps z_2 < 1/2,\;\; z_1>z_2
\end{array}\right. 
\label{eq:omega}
\end{equation}

We now turn to the calculation of the contact densities, which are the
order parameters of the OTW model. From the definition of the walk
model it immediately follows that the probabilities $\langle
\hat{a}_i\rangle_n$ and $\langle \hat{b}_i\rangle_n$ to have a contact
at site $2i$ above or below the x-axis, are given by,
\begin{eqnarray}
\langle \hat{a}_i\rangle_n &=& (z_1z_2)^i\frac{\tilde{Z}_{i}(z_1,\infty)
Z_{n-i}(z_1,z_2)}{Z_{n}(z_1,z_2)}, \label{eq:n1exp}\\
\langle \hat{b}_i\rangle_n &=& (z_1z_2)^{n-i} \frac{Z_{i}(z_1,z_2)
\tilde{Z}_{n-i}(\infty,z_2)}{Z_{n}(z_1,z_2)}. \label{eq:n2exp}
\end{eqnarray}
We define the average number of contacts at $x$ by
\begin{equation}
\langle \hat{a}_x\rangle = \langle \hat{a}_{xn}\rangle_n,
\qquad 
\langle \hat{b}_x\rangle = \langle
\hat{b}_{xn}\rangle_n,
\end{equation}
and find in the thermodynamic limit $n\rightarrow\infty$,
\begin{equation}
(\langle \hat{a}_x\rangle,\langle \hat{b}_x\rangle) = \left\{
\renewcommand{\arraystretch}{1.8}
\begin{array}{ll}
(0,0) & \quad \dps z_1,z_2 \geq 1/2 \\
\dps \left( \rho(z_1),0\right) & \quad\dps z_1 <1/2,\;\; z_2>z_1\\
\dps \left( 0,\rho(z_2)\right) & \quad\dps z_2 <1/2,\;\; z_1>z_2
\end{array}\right. 
\end{equation}
with 
\begin{equation}
\rho(z) = \frac{1-2z}{1-z}.
\label{eq:dens}
\end{equation}
In this limit, these numbers are independent of $x$ except on the
line $z_1=z_2=z$ where we find
\begin{equation}
(\langle \hat{a}_x\rangle,\langle \hat{b}_x\rangle) = \left(
\rho(z)(1-x),\rho(z) x\right).
\end{equation}
The total number of contacts above and below are denoted by $\langle
\hat{a}\rangle$ and $\langle \hat{b}\rangle$ respectively,
and the corresponding thermodynamic densities can be calculated
through derivatives of the grand potential,  
\begin{equation}
a = \lim_{n\rightarrow\infty} \frac{\langle \hat{a}\rangle}{n} 
= 1+ z_1 \frac{\partial \omega}{\partial z_1},\qquad b = 1+ z_2
\frac{\partial \omega}{\partial z_2}.
\label{eq:dens_der}
\end{equation}
Note that $\omega$ is not everywhere differentiable. If $\omega$ is
not differentiable in a point $z_*$ we define  
\begin{equation}
z_* \frac{\partial \omega}{\partial z}(z_*) = \lim_{\varepsilon \rightarrow 0}
\frac12 \left( (z_*-\varepsilon) \frac{\partial \omega}{\partial
z}(z_*-\varepsilon) +  (z_*+\varepsilon) \frac{\partial \omega}{\partial
z}(z_*+\varepsilon) \right).
\end{equation}
With this definition, (\ref{eq:dens_der}) is valid everywhere. Because
we average over the position of the domain wall, the densities $a$ and
$b$ are not independent. Their values can be easily calculated and are
given by,
\begin{equation}
(a,b) = \left\{
\renewcommand{\arraystretch}{1.8}
\begin{array}{ll}
(0,0) & \quad \dps z_1,z_2 \geq 1/2 \\
\dps \left( \rho(z_1),0\right) & \quad\dps z_1 < 1/2,\;\;z_2>z_1,\\
\dps \left( 0,\rho(z_2)\right) & \quad\dps z_2 < 1/2,\;\;z_1>z_2\\
\dps \left( \rho(z)/2,\rho(z)/2\right) & \quad\dps z_1=z_2=z \leq 1/2.
\end{array}\right. 
\label{eq:dens_ann}
\end{equation}
Notice that either both densities are equal or only one of the two
does not vanish. This means that effectively one sees only one density.
We thus find that for $z_1,z_2 >1/2$ the walk is entirely desorbed from the
interface. When $z_1<1/2$ and $z_2>z_1$ the walk is adsorbed above the
interface, the contact density $a$ is nonzero, while it is desorbed
below and vice versa when $z_2<1/2$ and $z_1>z_2$, see Figure
\ref{fig:phase}.

\begin{figure}[h]
\centerline{
\begin{picture}(215,210)
\put(10,10){\epsfxsize=200pt\epsfbox{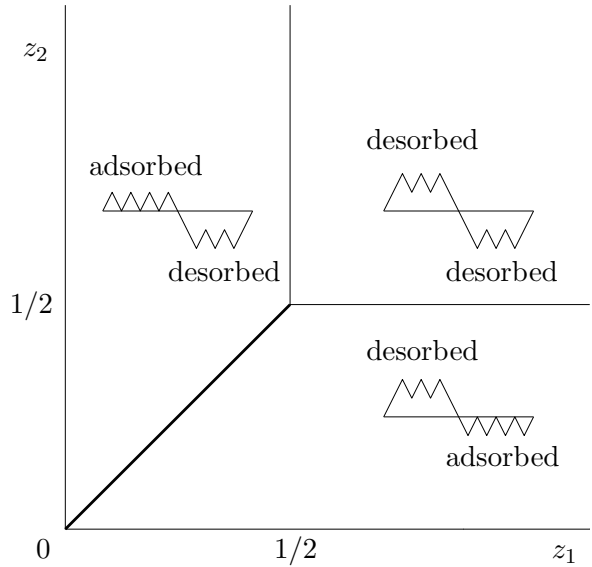}}
\put(0,0){$0$}
\put(90,0){$1/2$}
\put(-10,93){$1/2$}
\put(195,0){$z_1$}
\put(-5,190){$z_2$}
\put(125,155){desorbed}
\put(155,105){desorbed}
\put(125,75){desorbed}
\put(155,35){adsorbed}
\put(20,145){adsorbed}
\put(50,105){desorbed}
\end{picture}}
\caption{Phase diagram of the walk model.}
\label{fig:phase}
\end{figure}

The grand potential (\ref{eq:omega}) is non-analytic at the lines
$z_1=1/2$ when $z_2\geq z_1$ and $z_2=1/2$ when $z_1\geq z_2$. There
is also a singularity at the line $z_1=z_2=z$ when $z<1/2$. These
lines therefore indicate phase boundaries. 
There is a first-order phase transition along the line $z_1 = z_2=z$
for $z< 1/2$ along which the mirror symmetry of the system is
spontaneously broken. The sum of the two densities $r = a + b$ varies 
continuously across the line but their difference $d = a - b$ is discontinuous.
 
Above the line $z_2=1/2$ ($z_1\geq z_2$) the densities of contact points
$a= \langle \hat{a}\rangle /n$ and $b = \langle \hat{b}\rangle/n$
vanish as $n\rightarrow \infty$. Approaching the line $z_2 = 1/2$ from
above, the number of contacts $\langle \hat{b}\rangle$  diverges like
\begin{equation}
\langle \hat{b}\rangle \sim \frac{1}{2z_2-1},
\label{eq:above}
\end{equation}
where we have used (\ref{eq:asymp}). Using the same equation, on the
critical line $z_2 = 1/2$ we get
\begin{equation}
b \sim n^{-1/2} = n^{- \phi},
\label{eq:on}
\end{equation}
and below the critical line
\begin{equation}
b \sim 1 - 2z_2 = (1 -2z_2)^{1/\phi-1}.
\label{eq:below}
\end{equation}
A similar behaviour is obtained for the line $z_1=1/2$ when $z_2 \geq
z_1$ and if we replace the density $b$ by $a$. The critical behavior
(\ref{eq:above}) and (\ref{eq:below}) as well as the finite-size
scaling behavior (\ref{eq:on}) characterize a special surface phase
transition \cite{Binder83} with a single critical exponent $\phi =
1/2$. A similar exponent is found in other equilibrium problems with
long range interactions \cite{BrakEO98}. The interest in discussing
the phase diagram comes from the fact that it gives another physical
interpretation of the phase transitions observed in TASEP.

We conclude this section with a description of the model using the
canonical ensemble. We now
consider $a$ and $b$ as free parameters. The canonical free energy per
site for given values of $a$ and $b$ can be calculated from the grand
potential $\omega(z_1,z_2)$,  
\begin{equation}
f(a,b) = \sup_{z_1,z_2}( (1-a)\log z_1 +
(1-b) \log z_2+\omega(z_1,z_2)),
\label{eq:freedef}
\end{equation}
from which we find
\begin{eqnarray}
f(a,b) &=& \max\{ g(a,b),g(b,a) \},\\
g(a,b) &=& (1-a-b)\log(1-a)-(2-a-b)\log(2-a).
\end{eqnarray}
This result can be conveniently rewritten using $r=a+b$ and $d=a-b$ as
\begin{equation}
f(a,b) = (1-r)\log\left(1-\frac{r+|d|}{2}\right) - (2-r)\log\left(2-\frac{r+|d|}{2}\right).
\end{equation}

\subsection{Connection with the totally asymmetric
simple exclusion process}
\label{se:TASEP}

In \cite{DerrEHP93} the stationary state of the TASEP was constructed
using equivalent representations of the DEHP algebra. The reduced
normalisation calculated using this method is equal to the partition
sum of the OTW model, given by (\ref{eq:partsum1}), if the boundary
rates of the TASEP are replaced by the contact fugacitites of the OTW
model. In this section we show more precisely how the OTW model of
Section \ref{se:walk} is related to the TASEP.  

Following an observation by Brak and Essam \cite{BrakE03} (see also
see also \cite{DerrEL03}) that the different equivalent
representations of the DEHP algebra can be interpreted as 
transfer matrices for various lattice walk models, we construct a new
representation which will give the transfer matrix for the OTW model. 

\subsubsection{Transfer matrix formalism}
The OTW model can be described using a transfer matrix formalism. We
will show that the partition function $\tilde{Z}_n(z_1,z_2)$ can be written in
the following form
\begin{equation}
\tilde{Z} = \bra L T^n\ket R,
\label{eq:partsum3}
\end{equation}
where T is the transfer matrix. We introduce a two-step transfer
matrix $T=T^{\rm o} T^{\rm e}$, where 
\begin{equation}
T^{\rm o} = \left(
\begin{array}{@{}cc@{}}
D_1 & S \\
0 & D_2
\end{array}\right),\qquad
T^{\rm e} = \left(
\begin{array}{@{}cc@{}}
E_1 & 0 \\
0 & E_2
\end{array}\right).
\end{equation}
The matrices $D_1$ and $E_1$ will act as transfer matrices for the
walk above the x-axis, and $D_2$ and $E_2$ for the walk below the
x-axis. The upper triangular form of $T^{\rm o}$ ensures then that the
walk can cross the x-axis only once. We will now describe the transfer
matrices in detail.

The matrix element $(D_1)_{ij}$ for $j\geq 2$ is the weight of an edge
from a point with height $y=2i-2$ to a point with height $2j-3$. The
first column of $D_1$ is auxiliary whose meaning will become clear
later. Similarly, $(D_2)_{ij}$ for $j\geq 
2$ denotes the weight of an edge from a point with height $y=2-2i$ to
a point with height $3-2j$ and the first column is again auxiliary. If
the ket $\ket{n}$  represents the height $n$, the matrices $D_1$,
$D_2$ and $S$ are given in terms of projectors as,
\begin{eqnarray}
D_1 &=& \sum_{n=0}^{\infty}
\left(\ket{2n}+\ket{2n+2}\right)\bra{2n+1},\\
D_2 &=& x_1 \ket{0}\bra{{u}}+ \sum_{n=0}^{\infty}
\left(\ket{{-2n}}+\ket{{-2n-2}}\right)\bra{{-2n-1}},\\
S &=& x_2 \ket{0}\bra{{u}} + \ket{0}\bra{{-1}},
\end{eqnarray}
where $\ket{u}$ denotes an auxiliary ket vector. Explicitly, the
matrices $D_1$ and $D_2$ are given by,
\begin{equation}
D_1 = \left(
\begin{array}{@{}ccccc@{}}
0 & 1 & 0 & 0 & 0 \\
0 & 1 & 1 & 0 & 0 \\
0 & 0 & 1 & 1 & 0 \\
0 & 0 & 0 & 1 & \smash{\ddots} \\
0 & 0 & 0 & 0 & \smash{\ddots}
\end{array}\right),\qquad
D_2 = \left(
\begin{array}{@{}ccccc@{}}
x_1 & 1 & 0 & 0 & 0 \\
0 & 1 & 1 & 0 & 0 \\
0 & 0 & 1 & 1 & 0 \\
0 & 0 & 0 & 1 & \smash{\ddots} \\
0 & 0 & 0 & 0 & \smash{\ddots}
\end{array}\right),
\end{equation}
and the matrix $S$ is given by,
\begin{equation}
S = \left(
\begin{array}{@{}ccccc@{}}
x_2 & 1 & 0 & 0 & 0 \\
0 & 0 & 0 & 0 & 0 \\
0 & 0 & 0 & 0 & 0 \\
0 & 0 & 0 & 0 & \smash{\ddots} \\
0 & 0 & 0 & 0 & \smash{\ddots}
\end{array}\right).
\end{equation}
The parameters $x_1$ and $x_2$ are arbitrary and will not enter the
partition sum. For simplicity we could therefore set them to zero, but
we will need them later for another reason (see equation (\ref{eq:specx})).

The matrix element $(E_1)_{ij}$ for $i\geq 2$ is the weight of an edge
from a point with height $y=2i-3$ to a point with height $2j-2$. The
first row of $E_1$ is auxiliary. Similarly, $(E_2)_{ij}$ for $i\geq 2$
denotes the weight of an edge from a point with height $y=3-2i$ to a
point with height $2-2j$ and its first row is again auxiliary. In
terms of projectors, the matrices $E_1$ and $E_2$ are given by,
\begin{eqnarray}
E_1 &=& x_3 \ket{u}\bra{0} + z_1^{-1} \ket{{1}}\bra{0} + \sum_{n=1}^{\infty}
\left(\ket{2n-1}+\ket{2n+1}\right)\bra{2n},\\
E_2 &=& z_2^{-1} \ket{{-1}}\bra{0} + \sum_{n=1}^{\infty}
\left(\ket{{-2n-1}}+\ket{{-2n+1}}\right)\bra{-2n}.
\end{eqnarray}
Explicitly, they are given by,
\begin{equation}
E_1 = \left(
\begin{array}{@{}ccccc@{}}
x_3 & 0 & 0 & 0 & 0 \\
z_1^{-1} & 1 & 0 & 0 & 0 \\
0 & 1 & 1 & 0 & 0 \\
0 & 0 & 1 & 1 & 0 \\
0 & 0 & 0 & \smash{\ddots} & \smash{\ddots}
\end{array}\right),\qquad
E_2 = \left(
\begin{array}{@{}ccccc@{}}
0 & 0 & 0 & 0 & 0 \\
z_2^{-1} & 1 & 0 & 0 & 0 \\
0 & 1 & 1 & 0 & 0 \\
0 & 0 & 1 & 1 & 0 \\
0 & 0 & 0 & \smash{\ddots} & \smash{\ddots}
\end{array}\right),
\end{equation}
Also here, the parameter $x_3$ is arbitrary and will not enter the
partition sum. 

To indicate that walks can only start and end at height
$0$ we furthermore define the vectors $\bra{L}=\left({}_1\bra{L},
{}_2\bra{L}\right)$ and $\ket{R}=\left(\ket{R}_1,\ket{R}_2\right)$,
such that  
\begin{equation}
\renewcommand{\arraystretch}{1.4}
\begin{array}{ll}
\dps{}_1\bra{L} = \bra{0} = (1,0,0,\dots), & \dps {}_2\bra{L} = 0,\\
\dps\ket{R}_1 = \ket{0} = (1,0,0,\dots), & \dps\ket{R}_2 = \ket{0} =
(1,0,0,\dots).
\end{array}
\end{equation}
It is straightforward to check that the partition
sum (\ref{eq:partsum1}) of all walks of length $2n$ can
be expressed as (\ref{eq:partsum3}).

We end this section by defining the even and odd identity matrices for
future convenience,
\begin{equation}
I^{\rm o,e} = \left(
\begin{array}{@{}cc@{}}
I^{\rm o,e}_1 & 0 \\
0 & I^{\rm o,e}_2
\end{array}\right),
\end{equation}
where
\begin{eqnarray}
I^{\rm o}_1 = \ket{0}\bra{u} + \sum_{n=1}^{\infty}
\ket{2n}\bra{2n-1}, \quad I^{\rm o}_2 = \ket{0}\bra{u} +
\sum_{n=1}^{\infty} \ket{{-2n}}\bra{{-2n+1}},\\
I^{\rm e}_1 = \ket{u}\bra{0} + \sum_{n=1}^{\infty}
\ket{2n-1}\bra{2n}, \quad I^{\rm e}_2 = \ket{u}\bra{0} +
\sum_{n=1}^{\infty} \ket{{-2n+1}}\bra{{-2n}}.
\end{eqnarray}

\subsubsection{The TASEP revisited}
\label{se:tasep} 

The asymmetric simple exclusion process in continuous time is a
particle hopping model with excluded volume in one dimension, where
particles hop from the left to the right with rate $1$. In the
presence of open boundaries, the input rate of particles on the left
of the system is $\alpha$ and the output rate on the right is $\beta$,
see Fig. \ref{fig:asep}.
\begin{figure}[h]
\centerline{
\begin{picture}(300,50)
\put(0,0){\epsfxsize=300pt\epsfbox{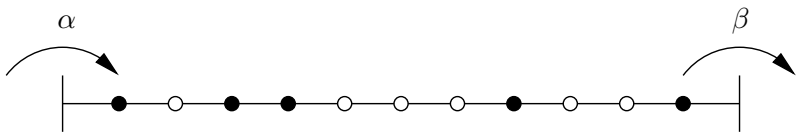}}
\put(20,40){$\alpha$}
\put(275,40){$\beta$}
\end{picture}}
\caption{Sample TASEP configuration. Particles enter the system from
the left with rate $\alpha$ and leave from the right with rate $\beta$.
Particles hop in the bulk from left to right with rate $1$.}
\label{fig:asep}
\end{figure}

If the $\tau_i \in \{0,1\}$ denotes the presence or absence of a
particle, one would for example like to know the probability
$P(\tau_1,\tau_2,\ldots,\tau_n\}$ to find a system in configuration
$\{\tau_1,\tau_2,\ldots,\tau_n\}$ in the long time limit. In this
limit, all these probabilities are stationary, and in \cite{DerrDM92}
this stationary state was calculated exactly. In \cite{DerrEHP93} it
was shown that this solution can be conveniently expressed in a matrix
product form,
\begin{equation}
P(\tau_1,\ldots,\tau_n) = \frac{1}{\tilde{Z}_n} \langle W| \prod_{i=1}^n (\tau_i D +
(1-\tau_i)E) |V\rangle,
\label{eq:mp}
\end{equation}
where the normalisation $\tilde{Z}_n$ is given by 
\begin{equation}
\tilde{Z}_n =\langle W|(D+E)^n |V\rangle,
\label{eq:ASEPnorm}
\end{equation}
and the matrices $D$ and $E$, and the vectors $\langle W|$
and $|V\rangle$ are a representation of the so-called DEHP algebra, 
\begin{eqnarray}
&&DE = D+E \nonumber\\
&&D |V\rangle = \frac1\beta |V\rangle 
\label{eq:dehp}\\
&&\langle W|E = \frac1\alpha \langle W|.\nonumber
\end{eqnarray}

From the result (\ref{eq:partsum1}) one may already have
inferred that the partition sum of the OTW model is equal to the
normalisation of the stationary state of the asymmetric simple exclusion
process (ASEP) with open boundaries \cite{DerrDM92}. The fugacities
$z_1$ and $z_2$ are then identified with the in- and output rates
$\alpha$ and $\beta$ respectively. Indeed, if we set the
parameters $x_i$ to the values,
\begin{equation}
x_1 = z_2^{-1},\qquad x_2=z_2^{-1},\qquad x_3=z_1^{-1},
\label{eq:specx}
\end{equation}
we find that the transfer matrices $T^{\rm o}$ and $T^{\rm e}$, and the vectors
$\bra{L}$ and $\ket{R}$ constitute the following representation of the DEHP
algebra (\ref{eq:dehp}), 
\begin{equation}
D = T^{\rm o} I^{\rm e},\qquad E=I^{\rm o} T^{\rm e}, \qquad
|V\rangle= \ket{R},\qquad \bra{W}=\bra{L}.
\end{equation}
Various representations of \cite{DerrEHP93} for the DEHP matrices
were used as transfer matrices in \cite{BrakE03} to find bijections
between several different path problems. Here we remark that this
interpretation of the DEHP matrices allows to express a stationary
nonequilibrium probability distribution in terms of an equilibrium
distribution. Among other things, this has the consequence that the
thermodynamics of the nonequilibrium model is prescribed by standard
equilibrium thermodynamics. 

The TASEP can be formulated using a transition matrix, see
e.g. \cite{DerrEHP93}. The normalisation as defined by
(\ref{eq:ASEPnorm}) is not equal to that calculated from 
the cofactors of this transition matrix using the results of Section
\ref{se:matrixtree}. In this approach, for each system size, all the
cofactors have a common factor which is a nontrivial
polynomial. As hinted at in Section \ref{se:matrixtree}, we believe
that this common factor will not give rise to additional singularities
for positive real rates in the thermodynamic limit. For example, upon
introduction of inhomogeneities in the transition rates, the cofactors
will no longer have a common factor. However, if these inhomogeneities are
small enough the physical properties of the system should remain the 
same. Moreover, we believe that a common factor is nonzero in the
space of complex rates if the real parts of all the rates are
positive. This is called the half-plane property, see
e.g. \cite{ChoeOSW02}. We have checked this for small system sizes in
the case of the TASEP. In the spirit of the Lee-Yang theory it 
implies that the common factor does not develop singularities in the
thermodynamic limit for positive real rates and hence it will not
influence the phase diagram, except perhaps at the origin.  

\subsection{OTW-TASEP relation}
\label{se:rel}
In this section we show how the TASEP current and density can be
related to the equilibrium densities of the OTW model.

\subsubsection{Current}
The TASEP current operator is given by,
\begin{equation} 
\hat{J} = (T^{\rm o} I^{\rm e})(I^{\rm o} T^{\rm e}).
\label{eq:currop}
\end{equation}
The average value,
\begin{equation}
J_{n,i} = \langle \hat{J}_i \rangle_n = \frac{1}{\tilde{Z}_n}\bra{L} T^{i-1}
\hat{J} T^{n-i-1}\ket{R}_n,
\end{equation}
has the following meaning in the path problem. The two identity
matrices in (\ref{eq:currop}) have, above the x-axis, the effect of
forcing an upstep between column $2i-1$ and $2i$ and a downstep between
$2i$ and $2i+1$. Below the $x$ axis they have the effect of forcing a
downstep between column $2i-1$ and $2i$ and an upstep between $2i$ and
$2i+1$. Therefore, $J_i$ is the average number of paths that have a 
local maximum above or a local minimum below the x-axis between
columns $2i-1$ and $2i+1$. The pieces of the path before and after
these local extrema can be concatenated to obtain a path of length $2n-2$. Since
the local extrema may occur at any height, we thus obtain all paths of
length $2n-2$ and therefore
\begin{equation}
J_{n,i} = J_n = \frac{\tilde{Z}_{n-1}}{\tilde{Z}_{n}},
\end{equation}
independent of $i$. In the OTW model, the current corresponds to the
pressure, since it is essentially the volume derivative of the grand
potential. The value of the current
\begin{equation}
J=\lim_{n\rightarrow\infty} J_n,
\end{equation}
in the various parts of the phase diagram is 
\begin{equation}
J = \left\{
\renewcommand{\arraystretch}{1.8}
\begin{array}{ll}
\dps 1/4  & \quad\dps z_1,z_2 \geq 1/2 \\
\dps z_1(1-z_1) & \quad z_1< 1/2,\;z_2>z_1\\
\dps z_2(1-z_2) & \quad z_2< 1/2,\; z_1>z_2
\end{array}\right.
\end{equation}

\subsubsection{Density}
The contact operators can be given in terms of projectors,
\begin{equation}
\hat{a}_i = \ket{1}_{2i-1} \bra{0}_{2i}, \qquad \hat{b}_i =
\ket{{-1}}_{2i-1} \bra{0}_{2i},
\end{equation} 
so that the contact number operators can be rewritten as
\begin{equation}
\hat{a} = \sum_{i=1}^n \hat{a}_i,\qquad \hat{b}  = \sum_{i=1}^n \hat{b}_i.
\end{equation}
The TASEP density operator $\hat{\tau}_i$ also has an expression in terms of the
projectors of the OTW model. The operator $\hat{\tau}_i$ is obtained by
putting the matrix $I^{\rm e}$ instead of $T^{\rm e}$ at position
$2i$
\begin{equation}
\hat{\tau}_i = I^{\rm e}(2i).
\end{equation}
This has the effect that between columns $2i-1$ and $2i$ each 
walk above the x-axis must go up. Walks below the x-axis must go down
between these columns at all heights except $y=-1$, where it also may
go up.
From the result of Derrida et al. \cite{DerrEHP93}, or from a
combinatorial argument \cite{BrakE03} it follows that the expectation
value $\langle \hat{\tau}_i\rangle$ can be written as 
\begin{eqnarray}
\langle \hat{\tau}_i\rangle_n &=& \frac{1}{\tilde{Z}_{n}(z_1,z_2)} 
\left[ \sum_{p=0}^{n-i-1} C_p \tilde{Z}_{n-p-1}(z_1,z_2) + \frac{1}{z_2}
  \tilde{Z}_{i-1}(z_1,z_2) \tilde{Z}_{n-i}(\infty,z_2) \right],
\end{eqnarray}
where we have used (\ref{eq:catalan}) and (\ref{eq:partsum1}).
Using the expression for the expectation values of the contacts
$\langle\hat{a}_i\rangle_n$ and $\langle\hat{b}_i\rangle_n$, see
eqs. (\ref{eq:n1exp}) and  (\ref{eq:n2exp}), we find
\begin{equation}
\langle\hat{\tau}_i \rangle_n = \sum_{p=0}^{n-i-1} C_p \prod_{j=0}^p
J_{n-j} + \frac{1}{z_2} J_{n-1} \langle \hat{b}_{i-1}\rangle_{n-1},
\label{eq:asep-walk_dens}
\end{equation}
and with the particle-hole symmetry of the TASEP this is equivalent to
\begin{equation}
\langle \hat{\tau}_i\rangle_n = 1- \sum_{p=0}^{i-2} C_p \prod_{j=0}^p
J_{n-j} - \frac{1}{z_1} J_{n-1} \langle \hat{a}_{i-1}\rangle_{n-1}.
\label{eq:asep-walk_dens2}
\end{equation}
Equations (\ref{eq:asep-walk_dens}) and  (\ref{eq:asep-walk_dens2})
give the relations between the local densities of the OTW model and
the TASEP. Combining (\ref{eq:asep-walk_dens}) and
(\ref{eq:asep-walk_dens2}) we find 
\begin{equation}
\langle \hat{\tau}_i\rangle_n = \frac12\left[ 1+ \sum_{p=i-1}^{n-i-1} C_p \prod_{j=0}^p
J_{n-j} + J_{n-1}\left( \frac{1}{z_2}\langle \hat{b}_{i-1}\rangle_{n-1} -\frac{1}{z_1} \langle
\hat{a}_{i-1}\rangle_{n-1}\right)\right].
\label{eq:asep-walk_dens3}
\end{equation}
In the bulk, the second term in the right hand side of
(\ref{eq:asep-walk_dens3}) vanishes in the thermodynamic 
limit. We thus find that in each part of the phase diagram the
following relation between the TASEP bulk density $\rho$ and current
$J$, and the equilibrium densities $a$ and $b$ is satsified, 
\begin{equation}
\frac{2\rho-1}{J} = \frac{b}{z_2}-\frac{a}{z_1},
\end{equation}
where $z_1=\alpha$, $z_2=\beta$ and $a$ and $b$ are given by (\ref{eq:dens_ann}).

\subsection{The partially asymmetric exclusion process}
\label{se:pasep}

The TASEP can be extended with a nonzero rate $q$ for back-hopping. The
resulting model is called the partially asymmetric simple exclusion
process (PASEP). Exact results for the symmetric case ($q=1$) are
given in \cite{Sasamoto96} and the stationary state of the general
PASEP can be found using a matrix method as well
\cite{Sasamoto99,BlytheECE00}.  To the rate $q$ will now correspond
a ``number of particles'' $N(q)$, not present in the TASEP. For the
forward bias regime ($q<1$) the phase structure is similar to the
TASEP model, we are interested in a new phenomenon which occurs in the
vicinity of $q=1$. One finds 
\begin{equation}
\renewcommand{\arraystretch}{2.2}
\begin{array}{rcll}
\dps N(q)&=& \dps \frac{q}{1-q} n + O(\log n),\qquad & \dps q<1,\quad\alpha,\beta >
  (1-q)/2 \\
\dps N(q)&=& \dps\frac14 n^2 + O(n), & q>1
\end{array}
 \label{eq:N}
\end{equation}
This implies a change of the volume when the transition rate $q$
changes. We notice that the density $\rho(q)= N(q)/n$ defined for $q <
1$ diverges for a finite value of $q$, namely $q = 1$. For $q>1$ we
have to redefine the density as $\rho(q) =  N(q)/n^2$. This density
turns out to be independent of the fugacity. A
change of the volume was observed also in TASEP at the phase transition
between the disordered state and the maximum current state. The latter
phase transition could be interpreted as a special surface phase
transition known in polymer physics (the number of ``particles'' were either
proportional with the size of the system or independendent of the size of
the system). Equation (\ref{eq:N}) describes a different phase transition since
the number of ``particles'' are either proportional with the size of the
system or with the square of the size of the system. We expect therefore that a simple
extension of the OTW model could explain what one observes.
This is indeed the case. In the new model, a walk gets height dependent
step weights. One can formulate a thermodynamical theory, analogous to that
of a heterogeneous gas in a gravitational field \cite{Gibbs} and obtain a
partition function equal to the normalisation factor of \cite{BlytheECE00}.
For $q>1$ the OTW model is genuinely two-dimensional, and has
therefore a volume of order $n^2$. For $q<1$ the system undergoes a
bulk phase transition and the only contributions to the grand
potential now come from the surface, i.e. the system becomes
one-dimensional. As in the TASEP, for $q<1$ the system may undergo
further phase transitions through enhancement of the surface chemical
potentials. This is indeed the what happens and we find the
adsorption-desorption transitions discussed in Section
\ref{se:thermo}. 

\section{Conclusion}
\label{se:conc}

In a previous paper \cite{GierNPR03} we have shown that a properly
chosen normalisation factor of the probability distribution function
describing the stationary state of the ``raise and peel''
one-dimensional model is given by the partition function of the
two-dimensional ice model with domain-wall boundary conditions $-$ an
equilibrium problem with nonlocal interactions. This connection was
proven for small systems but there are good reasons \cite{PearceRGN02}
to believe that this conjecture is valid for any size of the system. 
It turns out that the same way to choose the normalisation factor was
suggested in a general framework by Blythe and Evans \cite{BlytheE03}
and shown to be useful in order to use the Lee-Yang approach to
non-equilibrium problems. This brought us to have a closer look at the
problem. We have first noticed that the ``number of particles''
associated with various rates are non-decreasing functions of the
rates seen as fugacities. This allows, as in equilibrium problems,
to determine directly the phase diagram of a model, once the
normalisation factor is known. This observation suggests, obviously,
that one can try approximative approaches like finite-size scaling or
power expansions to determine the nature of the phase transition in
the case when the normalisation factor is not known exactly for all
sizes.

We have also shown, in the example of TASEP that, as in the ``raise and
peel'' model,  the normalisation factor can be understood as a partition
function of a two-dimensional equilibrium model: the one-transit walk
model. We also think that the correspondence between normalisation factors
of one-dimensional stationary states and two-dimensional equilibrium
problems with nonlocal interactions is of a more general validity. 

\section*{Acknowledgement}
Our warm thanks go to Erel Levine and Martin Evans for reading the
manuscript and to David Mukamel for discussions. Financial support
from the Australian Research Council and the European Commission network
HPRN-CT-2002-00325 is gratefully acknowledged.

\appendix
\section{The normalisation as a homogeneous polynomial}
\label{ap:proof}

The normalisation of a stationary state of any stochastic (Markov)
process can always be interpreted as a polynomial in the rates with
positive coefficients, i.e. it has the form of a generating
function. By the Cauchy-Schwartz inequality, the negative logarithm of this
generating function is therefore convex and its derivatives with 
respect to the rates are proper ``particle'' numbers, i.e. the second
derivatives are positive.

The statement above is implied by the matrix-tree theorem
\cite{Sylvester55,Chen76,Chaiken82}. Here we show a simple proof which
can be found in a slighly different version in \cite{Seneta81}.
\begin{lemma}
Let $H$ be a matrix with off-diagonal elements $H_{ab} = -r_{ab}$ and
such that all columns add up to zero, $\sum_{a} H_{ab}=0$. Assume that
$H$ has a unique largest eigenvalue equal to $0$.
\begin{itemize}
\item[a)] The cofactors $X(a,b)$ are constant for each column,
i.e. they do not depend on $a$.
\item[b)] The eigenvector corresponding to the largest eigenvalue $0$
is a polynomial in the rates $r_{ab}$ with positive coefficients.
\end{itemize}
\label{le:eigpos}
\end{lemma}
Proof.

Let $H(a,b)$ be the matrix corresponding to $H$ with the $a$th row and
$b$th column removed. The cofactor $X(a,b)$ is then defined by,
\begin{equation}
X(a,b) = (-1)^{a+b} \det H(a,b).
\end{equation}
Using row operations that do not change the determinant, and because
of the special properties of $H$, it is not difficult to transform
$H(a,b)$ into $H(b,b)$, and hence $X(a,b) = X(b,b)$ for all $a$. 

Because the eigenvalue $0$ is unique and,
\begin{equation}
0 = \det H = \sum_b H_{ab} X(a,b) = \sum_b H_{ab} X(b,b),
\end{equation}
the elements of the eigenvector corresponding to the
eigenvalue $0$ are given by the cofactors $X(b,b)$.  Each such
cofactor is of the form
\begin{equation}
X(b,b) = \sum_{\pi} N(\pi_1,\ldots,\pi_{b-1},\pi_{b+1},\ldots,\pi_n)
\prod_{\stackrel{c=1}{\scriptscriptstyle c\neq
b}}^n r_{\pi_c,c}, 
\label{eq:cofac_gen}
\end{equation}
where the sum is over any permutation
$\pi=\{\pi_1,\ldots,\pi_{b-1},\pi_{b+1},\ldots,\pi_n\}$ of
$\{1,\ldots,b-1,b+1,\ldots,n\}$ and $N(\pi)\in {\mathbb Z}$. We now
show that in fact $N(\pi)\in \{0,1\}$, hence proving assertion $b)$ of
Lemma \ref{le:eigpos}.  

Let $r_{\rho_a,a}=1$ for a particular permutation $\rho$, and all
other $r_{ac}=0$. From (\ref{eq:cofac_gen}) we then see that
$N(\rho_1,\ldots,\rho_{b-1},\rho_{b+1},\ldots, \rho_n)$ is the
determinant of a matrix which we will call $H(b,b,\rho)$. If $\rho_a
\neq b$ for all $a=1,\ldots,b-1,b+1,\ldots,n$, the columns of
$H(b,b,\rho)$ all add up to zero and $\det H(b,b,\rho)=0$. If on the
other hand $\rho_a=b$ for a particular $a=a^*$, then $H(b,b,\rho)$
contains zeros in the column corresponding to $a^*$ except for the
diagonal element which is $1$. By deleting the column and row of
$H(b,b,\rho)$ corresponding to $a^*$ we find again a matrix of the
form of $H(b,b,\rho)$ but with one dimension less. The result thus
follows by expanding the determinant with respect to the column
corresponding to $a^*$ and induction on $n$.

\end{document}